# ارایه‌ی مدل پویای تعاملات کارکردی نظام نوآورانه‌ی فناوری اینترنت اشیا با استفاده از پویایی‌های سیستم و دیماتل فازی

محمد موسی خانی*۱، فاطمه ثقفی۲، محمد حسن زاده۳، محمد ابراهیم صادقی۴

۱- دانشیار، دانشگاه تهران، دانشکده‌ی مدیریت، گروه مدیریت فناوری اطلاعات، تهران، ایران

۲- دانشیار، دانشگاه تهران، دانشکده‌ی مدیریت، گروه مدیریت صنعتی، تهران، ایران

۳- استاد، دانشگاه تربیت مدرس، دانشکده‌ی مدیریت و اقتصاد، گروه علم اطلاعات و دانش شناسی، تهران، ایران

۴- دانشجوی دکتری، دانشگاه تهران، دانشکده‌ی مدیریت، گروه مدیریت صنعتی، تهران، ایران

## چکیده

یکی از فناوری‌های نوظهوری که پیش‌بینی می‌شود تاثیرات شگرفی بر توسعه‌ی جوامع داشته باشد، فناوری اینترنت اشیا می‌باشد. نظر به چشم‌اندازهای پیش روی این فناوری و تلاش کشور برای توسعه‌ی آن، ضروری است تا برای سیاستگذاری آن اقدام گردد. نظام نوآورانه فناوری یکی از مهم‌ترین رویکردهای پویا در سیاستگذاری فناوری‌های نوین می‌باشد. در این دیدگاه با تحلیل کارکردهای متنوعی که بر روی توسعه‌ی یک فناوری تاثیرگذار می‌باشند، مسیر مناسب پیشرفت فناوری تبیین می‌گردد. به همین دلیل بر اساس مطالعات پیشین این حوزه، ۱۰ عامل اصلی تاثیرگذار بر توسعه‌ی فناوری‌های نوظهور شناسایی گردید و با استفاده از روش پویایی‌های سیستم و دیماتل فازی نحوه‌ی اثرگذاری عوامل دهگانه بر یکدیگر شناسایی شد و تعاملات میان این کارکردها مدل گردید. ۴ عامل شکل‌گیری بازار، بسیج منابع، بهره‌برداری از رژیم موجود و سیاستگذاری و ایجاد هماهنگی بیش‌ترین تاثیر مستقیم را بر سایر عوامل دارا بودند. همچنین ۵ عامل سیاستگذاری و ایجاد هماهنگی، شکل‌گیری بازار، فعالیت‌های کارآفرینانه، ایجاد ساختار و بسیج منابع دارای بیش‌ترین تاثیر به صورت کلی بودند. با توجه به محدودیت منابع در یک سیستم، باید عمده‌ی توجه سیاستگذار بر عواملی باشد که بیش‌ترین تاثیر مستقیم و کل و نیز بیش‌ترین روابط و حلقه‌های بازخوردی را دارا می‌باشند که در این تحقیق چهار عامل شکل‌گیری بازار، بسیج منابع، فعالیت‌های کارآفرینانه و سیاستگذاری و ایجاد هماهنگی می‌باشند. با توجه به ماهیت پویای توسعه‌ی فناوری، این مدل می‌تواند در فرایند تصمیم‌گیری برای توسعه‌ی اینترنت اشیا به سیاستگذاران کشور یاری رساند.

**کلمات کلیدی:** نظام نوآوری فناورانه، سیاستگذاری، اینترنت اشیا، پویایی سیستم، دیماتل فازی.

---

* عهده‌دار مکاتبات

آدرس الکترونیکی: muosakhani@ut.ac.ir

# ۱ مقدمه

پیشرفت گسترده‌ی علم و فناوری سبب تغییرات بزرگی در جوامع شده‌است. ظهور فناوری‌های جدید که مسیرهای جدیدی را در حل مسایل باز می‌نمایند به یکی از واقعیات جامعه‌ی امروزی تبدیل شده‌است. بنابراین دولت‌ها تمایل دارند تا از طریق سیاست‌گذاری، از این فناوری‌ها بهره‌مند شوند. دو رویکرد عمده در این حوزه وجود دارد. دیدگاه قدیمی‌تر در این حوزه، بهره‌گیری از تجمیع جغرافیایی در قالب خوشه‌های صنعتی و پارک‌های علم و فناوری در توسعه‌ی فناوری‌های پیشرفته بوده است [۱]، که همچنان حوزه‌ی مطالعاتی مهم و در حال رشدی می‌باشد. رویکرد جدیدتر در توسعه‌ی فناوری‌های نوین دیدگاه نظام نوآوری می‌باشد، که خود دارای شاخه‌های مختلفی مانند نظام نوآوری منطقه‌ای [۲]، نظام نوآوری ملی، نظام نوآوری بخشی و نظام نوآوری فناورانه [۳] می‌باشد. نظام نوآوری فناورانه، به عنوان یکی از حوزه‌های مطالعاتی در حوزه‌ی مطالعات نوآوری که به دنبال سیاست‌گذاری فناوری‌های نوین می‌باشد، در سال‌های اخیر رشد قابل توجهی داشته و مورد توجه محققین قرار گرفته‌است.

مداخله‌ی سیاستی به منظور توسعه‌ی فناوری‌های نوین یکی از مهم‌ترین دستاوردهای این رویکرد بوده‌است. خصوصا، در یک دهه‌ی گذشته چارچوب تحلیلی کارکردی در نظام نوآوری فناورانه به عنوان رویکرد غالب در تحلیل یک فناوری و ارایه‌ی پیشنهادات سیاستی مورد توجه قرار گرفته‌است. در اولین گام‌ها هکرت و همکاران (۲۰۰۷) و نگرو و همکاران (۲۰۰۷) و سپس برگک و همکاران (۲۰۰۸) از این رویکرد در نظام نوآوری فناورانه استفاده نمودند که پس از آنها استفاده از این چارچوب تحلیلی بسیار مورد توجه واقع شد و مورد استفاده قرار گرفت [۳-۵]. در این مدت محققین در مطالعات خود کارکردهای نظام نوآوری فناورانه را به لحاظ تعدادی و محتوایی غنا بخشیده‌اند و کارکردها و زیر کارکردهای متعددی را که بر شکل‌گیری و توسعه‌ی یک فناوری اثرگذار می‌باشند، بسط و توسعه داده‌اند. همچنین تحلیل‌های مبتنی بر ساختار و تحلیل‌های مبتنی بر ارتباط میان ساختار، فرآیندها و زمینه، فرصت‌های جدید مداخلات سیاستی را برای دولت‌ها فراهم نموده‌است. همان‌گونه که جاکوبسون و برگک (۲۰۱۱) بیان کرده‌اند تحلیل‌های مبتنی بر نظام نوآوری، ابزاری است که امکان شناسایی فرصت‌های مداخله در سطح سیستمی را فراهم می‌نماید [۶].

یکی از فناوری‌های نوظهوری که به نظر می‌رسد در سال‌های آینده به عنوان یک فناوری عام عمل خواهد نمود و تاثیر بسیار بالایی بر بهره‌وری اقتصاد و صنایع خواهد گذاشت، فناوری اینترنت اشیا می‌باشد. حرکت شتابان دولت‌ها و شرکت‌ها برای عملیاتی‌سازی پروژه‌های اینترنت اشیا نشان از آینده‌ی نویدبخش این فناوری دارد.

اینترنت اشیا یک پارادایم نوظهور برای اتصال تجهیزات فیزیکی، با هدف یکپارچه‌سازی بی‌واسطه‌ی دنیای فیزیکی به سیستم‌های کامپیوتری است که منجر به کارایی بیشتر، کاربردهای جدید و رشد اقتصادی می‌شود [۷]. چندین شی فیزیکی می‌توانند در اینترنت اشیا با یکدیگر مرتبط شده و داده‌ها را بر بستر اینترنت تبادل کرده و جمع‌آوری نمایند. اینترنت اشیا حوزه‌های کاربردی متنوعی دارد، حوزه‌هایی مانند پایش هوشمند ترافیک، خانه هوشمند، تجهیزات پوشیدنی، صنایع و شهر هوشمند. در یک محیط اینترنت اشیا ابری، از پلتفرم‌های ابری برای

۲

ذخیره‌سازی داده‌های سنسورهای اینترنت اشیا استفاده می‌شود. این محیط به شدت قابلیت مقیاس‌پذیری دارد و پردازش بلادرنگ رخدادها را میسر می‌سازد که در برخی از شرایط (برای مثال در کاربردهای پایش و مراقبت) جنبه‌ی حیاتی دارد. بنابراین، کاربردهای مبتنی بر اینترنت اشیا به یکی از بخش‌های ضروری زندگی روزمره‌ی ما تبدیل می‌شود [۸].

اینترنت اشیا یکی از پویاترین حوزه‌ها را در چشم‌انداز فناوری اطلاعات ارایه کرده‌است. علت آن نیز یکپارچه‌سازی فناوری‌های بسیار ناهمگون و نیز ظهور کاربردهای جدید مبتنی بر فناوری اینترنت اشیا در حوزه‌های مختلف است. حوزه‌هایی شامل شهر هوشمند، ساختمان‌های هوشمند، سیستم‌های سلامت الکترونیک، سیستم‌های ساخت و تولید هوشمند و سیستم‌های حمل و نقل هوشمند [۹].

ظهور اینترنت اشیا الگوی تفکر سنتی را ارتقا می‌دهد و امکان اتصال بسیاری از اشیا موجود در محیط -اگر نگوییم همه- را در چارچوب شبکه فراهم می‌آورد. این فناوری، خودروها، لوازم خانگی و سایر تجهیزات الکترونیکی را بر روی شبکه به یکدیگر متصل می‌نماید که در نتیجه برای بشر زندگی هوشمندانه‌تری را به ارمغان می‌آورد. سیستم امکان شناسایی، موقعیت‌یابی، رهگیری و نظارت بلادرنگ خواهد داشت و به صورت خودکار در برابر وقایع پاسخ خواهد داد. [۱۰].

در کمتر از یک دهه از مطرح شدن این مفهوم، در سال ۲۰۰۸ تعداد اشیایی که باهم مرتبط بودند از جمعیت کره زمین فراتر رفت و پیش‌بینی می‌گردد که تا سال ۲۰۲۰ تعداد آن‌ها حداقل به ۵۰ میلیارد و حداکثر به حدود ۱۰۰ میلیارد شی متصل‌به‌هم برسد [۱۱]. در آینده‌ی نزدیک، اینترنت اشیا در زندگی روزمره‌ی ما رسوخ خواهد کرد. بر اساس پیش‌بینی سیسکو سیستمز[1] تا سال ۲۰۲۰ نزدیک به ۵۰ میلیارد تجهیز دارای آی پی در اینترنت خواهیم داشت که به معنای یک فرصت ۱۴/۴ تریلیون دلاری در کسب و کار می‌باشد. پیش‌بینی می‌شود که همه‌ی مشاغل از اینترنت اشیا تاثیر بپذیرند [۱۲].

توسعه‌ی این فناوری در کشور مورد توجه سیاستگذاران حوزه‌ی فناوری‌های پیشرفته قرار گرفته‌است و تلاش‌هایی برای تدوین استراژی و نقشه‌ی راه برای آن صورت پذیرفته‌است، برای مثال می توان به پروژه «تدوین نقشه‌ی راه فناوری اینترنت اشیا» توسط پژوهشگاه ارتباطات و فناوری اطلاعات اشاره نمود. در همین راستا، معاونت علمی و فناوری ریاست جمهوری نیز اقدام به تهیه‌ی فهرست محصولات دانش‌بنیان در حوزه‌ی فناوری اینترنت اشیا نموده است. همچنین می توان به اسناد کلان تصویب شده در رابطه با فناوری اینترنت اشیا در شورای عالی فضای مجازی شامل «الزامات حاکم بر اینترنت اشیا در شبکه ی ملی اطلاعات» و «تقسیم کار ملی توسعه‌ی اینترنت اشیا در شبکه‌ی ملی اطلاعات» اشاره نمود. این موارد نشان‌دهنده‌ی اهمیت این فناوری در آینده‌ی کشور است.

---

[1] Cisco Systems

۳

## ۲ پیشینه تحقیق

نظام نوآوری فناورانه به عنوان یکی از زیرشاخه‌های رویکرد نظام‌های نوآوری مطرح شد. اولین دیدگاهی که در این رویکرد مطرح شد نظام ملی نوآوری بود. نظام‌های نوآوری اهمیت تعامل‌های «نظام‌مند» میان مولفه‌های مختلف اختراعات، تحقیق، تغییر فنی، یادگیری و نوآوری را به طور کاملا روشن بیان می‌کنند. همچنین، نظام‌های ملی نوآوری نقش محوری دولت را به عنوان عامل هماهنگ کننده برجسته می‌سازند. جذابیت ویژه این رویکرد برای سیاستگذاران در بازشناسی لزوم وجود سیاست‌های مکمل، توجه به ضعف‌های موجود در سیستم و در عین حال تبیین جایگاه ملی اغلب این نهادها بود [۱۳].

پس از آن رویکردهای دیگری همچون نظام بخشی نوآوری و نظام منطقه‌ای نوآوری نیز مطرح شدند [۱۴،۱۵]. جاکوبسون و برگک (۲۰۱۱) با بررسی مطالعات پیشین بیان می‌دارند که دیدگاه نظام نوآوری در اواسط دهه‌ی ۸۰ با هدف پاسخ به نارسایی‌های موجود در اقتصاد نئوکلاسیک و اشاعه‌ی نئولیبرالیزم در بحث‌هایی که پیرامون سیاست صنعتی در اروپا مطرح بود شکل گرفت. آنها بیان می‌دارند اگر چه نظام‌های نوآوری مختلف در مرزهای خود با یکدیگر اختلاف دارند، اما یک نظام نوآوری که بر یک فناوری خاص متمرکز شده‌است در زمینه‌ای از سطوح مختلف، ملی، منطقه‌ای و بخشی قرار گرفته‌است [۶].

در برخی مقالات، مقاله کارلسون و استنکویتز (۱۹۹۱) به عنوان اولین گام در معرفی مفهوم نظام فناورانه نوآوری مطرح می‌شود [۳،۵، ۱۶-۲۰]. اگر چه در آن مقاله صراحتا از عبارت نظام نوآوری فناورانه استفاده نشده‌است اما محققین بعدی از تعریف موجود در این مقاله در رابطه با نظام‌های فناورانه استفاده کرده‌اند. کارلسون و استنکویتز (۱۹۹۱) نظام‌های فناورانه را چنین تعریف نموده‌اند:

شبکه‌ای پویا از عوامل که در یک ناحیه‌ی اقتصادی/صنعتی تحت زیرساخت‌های نهادی خاص با یکدیگر در تعامل بوده و در تولید، انتشار و بهره‌برداری از فناوری سهیم هستند [۲۱].

دو چارچوب تحلیل عمده در نظام نوآوری فناورانه وجود دارد. یکی چارچوب تحلیل ساختاری و دیگری چارچوب تحلیل کارکردی. در مورد ارتباط کارکردها با ساختارها، برخی از محققین استدلال می‌کنند که کارکردها باید فرآیندهای ساختارساز قلمداد شوند (به طورمثال [۲۲، ۲۳]). از این منظر، کارکردها فرآیندهایی هستند که توسعه ساختارهایی نظیر بازیگران جدید، زیرساخت‌ها یا نهادها را شکل می‌دهند. برخی دیگر از محققین بیان می‌دارند که کارکردها ویژگی‌های درحال ظهور یک نظام نوآوری هستند که می‌توان از آنها در روش‌های تشخیصی استفاده نمود: «کارکردها، حالت یک نظام نوآوری خاص را در یک لحظه مشخص از زمان نشان می‌دهند» [۲۴]. برخی محققین معتقدند رویکرد کارکردی و ساختاری دو روی یک سکه هستند [۲۰] و رویکرد کارکردی به عنوان مرحله‌ی واسطی میان تحلیل ساختاری و سیاستگذاری برای سیاستگذاران عرضه شده‌اند. برخی دیگر از محققین، معتقدند که در هنگام تحلیل یک نظام نوآوری فناورانه باید به دو جنبه‌ی کارکردی و ساختاری سیستم توجه نمود و به منظور ترکیب آنها با یکدیگر پیشنهاداتی ارایه نموده و تلاش نموده‌اند تا هر کارکردسیستم را از طریق شکست‌های سیستمی به یک نارسایی ساختاری مرتبط نمایند [۲۵]. کاراناسیوس و پارکر (۲۰۱۸) بر اساس مطالعه‌ی وبر و روهراچر (۲۰۱۲) و وولتوئیس و همکاران (۲۰۰۵) بیان

۴

می‌دارند به منظور درک دقیق یک نظام نوآوری فناورانه باید یک تحلیل ترکیبی از ساختار و کارکردهای آن انجام داد تا از طریق شناسایی شکست‌های سیستمی بتوان برای توسعه‌ی آن چاره‌جویی نمود [۲۶-۲۸].

در مقالات متعددی از رویکرد نظام نوآوری فناورانه در کشورهای در حال توسعه استفاده شده است، بـرای مثال می‌توان به بررسی فناوری پیل سوختی در تایوان [۲۹]، بوم‌شناسی کشاورزی در نیکاراگوئـه [۳۰]، تجهیـزات ارتوپدی در آفریقای جنوبی [۳۱]، فناوری هضم بیولوژیکی در پاکستان [۳۲]، صـنایع ارتباطـات در چین [۳۳] و انرژی بادی در چین و برزیل [۳۴] اشاره نمود.

از این رویکرد در ایران نیز استفاده شده است که برای مثال می‌توان به بررسی صنایع پتروشیمی [۳۵]، نفت و گاز [۳۶]، هیدروژن و پیل سوختی [۳۷-۳۹] اشاره نمود.

تاکنون، محققان مختلف دسته‌بندی‌های متفاوتی از کارکردهای نظام‌های نوآوری ارایه نموده‌اند. برای اولین بار در تحقیقات حوزه نظام نوآوری توسط جانسون (۱۹۹۸) با معرفی ۸ کارکرد مجموعـه‌ای از کارکردهـا مطـرح شدند. پس از او ریکنـه (۲۰۰۰) نیز مجموعـه‌ای از کارکردهـا را مطرح نمـود، جانسـون (۲۰۰۱) نیـز مجموعه‌ی کارکردهای دیگری را منتشر نمود. محققان دیگری همچون ادکوئیست (۲۰۰۴)، کارلسون و همکاران (۲۰۰۵) و هکرت و همکاران (۲۰۰۷) نیز مجموعه‌ای از کارکردها را معرفی نمودند [۴]. هکرت و همکاران (۲۰۰۷) بـرای اولین بار رویکردی را تحت عنوان نظام نوآوری فناورانه استفاده نمودند کـه پـس از آنهـا اسـتفاده از ایـن چارچوب تحلیلی بسیار مورد توجه واقع شد و مورد استفاده قرار گرفت [۵]. همان‌طور که برگک و همکاران (۲۰۰۸) بیان می‌دارند، تعداد کارکردها تا حدود زیادی دلبخواهانه و مبتنی بر مطالعات پیشین محققان ایـن حوزه صورت پذیرفته‌است [۲۳].

با بررسی تحقیقات مختلف مشاهده می‌شود که موارد متعددی به عنوان کارکردهـای اصـلی نظام نـوآوری فناورانه مطرح شده‌اند. مواردی مانند توسعه‌ی دانـش، انتشار دانـش، فعالیت‌های کارآفرینانـه، جهـت دهـی بـه تحقیقات، شکل‌گیری بازار، مشروعیت‌بخشی و بسیج منابع که توسط هکرت و همکاران (۲۰۰۷) مطـرح شـده‌انـد [۵]، در بیشتر مطالعات مورد استفاده قرار گرفته‌اند. سایر محققین مواردی مانند توسعه‌ی اثرات خارجی مثبت و ساخت محصولات [۲۳]، تبادل دانش [۴۰]، تاثیرات سیاسی و اجتماعی سطح بالاتر [۴۱]، توسعه‌ی سـرمایه اجتماعی [۴۲]، ایجاد ظرفیت انطباقی [۴۳]، انتظارات و انتخاب‌های فناوری و واسطه‌گری علایق و لابی‌گری [۴۴] را اضافه نموده‌اند.

در این تحقیق بر اساس مطالعات پیشین این حوزه، با یکپارچه‌سازی و توسعه‌ی کارکردها تلاش شده‌است تا تمامی ابعاد لازم برای توسعه‌ی یک فناوری در نظر گرفته شود. در نتیجه ۱۰ عامـل اصـلی تاثیرگـذار بـر توسـعه‌ی فناوری‌های نوظهور شناسایی شده‌اند که در ادامه به توضیح مختصر هر یک از آنها می‌پردازیم.

توسعه، تبادل و انتشار دانش (F1):

دانش نقشی کلیدی در هر نظام نوآورانه دارد. ایـن کـارکرد را مـی‌توان به عنـوان سـوخت نظام نوآوری فناورانه در نظر گرفت [۴۵]. پس از خلق دانش، انتشار آن در کل سیستم به منظور بهره‌برداری حداکثری از آن بسیار مهم خواهد بود و به بلوغ سیستم کمک خواهد کرد. تبادل دانش خصوصا در سطح بین‌المللی امکان

۵

سرعت بخشیدن به توسعه‌ی فناوری را فراهم می‌آورد و با توجه به درهم‌تنیدگی هر چه بیشتر اقتصادها امکان جبران عقب ماندگی در فناوری‌ها را میسر می‌سازد.

فعالیت‌های کارآفرینانه (F2):

این کارکرد به بررسی حرکت یک نظام نوآورانه فناورانه به سوی پیاده‌سازی فناوری می‌پردازد و موفقیت‌ها در شکل‌گیری شرکت‌ها، محصولات و کاربردهای جدید را مدنظر قرار می‌دهد. هرچه فعالیت‌های کارآفرینانه بیشتر و موفق‌تر باشند امکان شکل‌گیری بازار و موفقیت فناوری به تبع آن بیشتر خواهد بود و مشروعیت فناوری در جامعه بیشتر خواهد شد. این کارکرد به دنبال یافتن کاربردها و بازارهای جدید برای فناوری است [۴۶].

جهت‌دهی به تحقیقات (F3):

این کارکرد به مجموعه اقداماتی که یک دیدگاه مشترک از آینده‌ی فناوری و توسعه‌ی آن را در میان آحاد جامعه ایجاد می‌نمایند، می‌پردازد. به بیان دیگر این کارکرد یک فرایند انتخاب که بر اساس آن راه‌حل‌های نوآورانه مورد ارزیابی قرار می‌گیرند و منجر به پذیرش فناوری در جامعه می‌شوند را در بر می‌گیرد [۲۶]. هرچند حضور دولت و تعیین اهداف فناورانه نقش مهمی در این زمینه ایفا می‌کند، اما از نقش مصرف‌کنندگان و خود صاحبان فناوری نباید غفلت نمود. با توجه به محدودیت منابع در هر سیستم این کارکرد جهت اصلی تحقیقات و حرکت سیستم در مسیر توسعه‌ی خود را تحت تاثیر قرار می‌دهد و مسیر حرکت را برای بازیگران مختلف مشخص می‌نماید.

شکل‌گیری بازار (F4):

تمرکز اصلی در این کارکرد ایجاد تقاضا برای فناوری تولید شده‌است. از آنجایی که فناوری‌های جدید در برابر فناوری‌های قدیمی در ابتدا از توان رقابتی پایین‌تری برخوردار هستند بنابراین باید تلاش نمود تا با ایجاد بازارهای گوشه، امکان افزایش سهم بازار را برای آنها فراهم نمود [۴۷]. در همین راستا تلاش می‌شود تا با اقدامات مختلف رقابت‌پذیری این فناوری در مقایسه با سایر فناوری‌ها بالا رفته و جذابیت مصرف آن افزایش یابد. هنگامی که فناوری از مرحله‌ی کودکی خود خارج می‌شود، این کارکرد مهم‌ترین کارکرد در توسعه‌ی فناوری خواهد بود و تا زمان افول فناوری بیش‌ترین نقش را برعهده خواهد داشت.

بسیج منابع (F5):

محدودیت منابع، مانع اصلی در عدم توفیق فناوری‌هاست. هرچه منابع بیشتری را بتوان به یک فناوری اختصاص داد احتمال موفقیت آن بیشتر خواهد بود. این کارکرد به عنوان توانمندساز عمل می‌نماید و پیاده‌سازی مناسب سایر کارکردها نیز وابسته به این کارکرد می‌باشد [۴۸]. صورت‌های مختلفی از منابع وجود دارند که در این کارکرد باید مورد توجه قرار گیرند. نقش دولت به عنوان فراهم‌کننده و یا تسهیل‌کننده در دسترسی به بسیاری از منابع کلیدی می‌باشد. با توسعه‌ی هرچه بیشتر فناوری، نقش بخش خصوصی در تامین منابع مختلف بیشتر می‌شود، اما کماکان نیاز به حضور موثر دولت در بسیج منابع خواهد بود.

مشروعیت‌بخشی (F6):

۶

وجود حامیان قدرتمند برای یک فناوری در درون دولت، بخش خصوصی و جامعه نقش مهمی در توسعه‌ی یک فناوری دارد. پذیرش فناوری در جامعه و تطابق آن با نهادهای مرتبط، نقشی حیاتی در موفقیت یک فناوری دارد [۴۹]. از آنجایی که یک فناوری نوظهور ممکن است منافع گروه‌های ذی‌نفوذ فعلی را به خطر اندازد، معمولا مخالفت‌هایی با آن در بخش‌های مختلف شکل می‌گیرد. لذا این کارکرد به تحکیم پایه‌های این فناوری در جامعه و در میان گروه‌های ذی‌نفوذ می‌پردازد.

سیاست‌گذاری و ایجاد هماهنگی (F7):

مهم‌ترین موضوع در این کارکرد، ایجاد یک ساختار سیاستی چابک و منعطف به منظور راهبری کل بازیگران در توسعه‌ی یک فناوری است. در این کارکرد تلاش می‌شود تا با ایجاد هماهنگی در بخش‌های مختلف و میان بازیگران مختلف نیروهای موجود در توسعه‌ی فناوری را هم‌افزا نمود و با مشارکت فعال همه‌ی بازیگران و با رعایت اصول سیاست‌گذاری خوب [۵۰]، مسیر توسعه‌ی فناوری را به طور مستمر بهبود بخشید.

ایجاد ساختار (شبکه و نهادهای واسط) (F8):

توسعه‌ی فناوری نیازمند شکل‌گیری ساختارهای متعددی می‌باشد. از مهم‌ترین این ساختارها می‌توان به شبکه‌ها و نهادهای واسط اشاره نمود. این نهادها از طریق ایجاد هم‌راستایی میان بازیگران مختلف نظام نوآوری شامل: سیاست‌گذاران، کارآفرینان، سرمایه‌گذاران، تولیدکنندگان و مصرف‌کنندگان نقش مهمی در توسعه‌ی یک نظام نوآورانه فناورانه ایفا می‌کنند [۵۱]. بنابراین شکل‌گیری و وجود شبکه‌ها و نهادهای واسطی که با ارایه‌ی خدمات تخصصی خود به فناوران زمینه‌ساز رشد و حضور موفق آنها در بازار می‌شوند یکی از مهم‌ترین ارکان پیشرفت فناوری‌های نوظهور می‌باشد.

تضعیف رژیم حاکم (F9):

همان‌طور که در کارکرد مشروعیت‌بخشی اشاره شد، معمولا رژیم حاکم سعی در مقابله با فناوری نوظهور دارد، از همین رو کارکرد مشروعیت‌بخشی به منظور ایجاد ائتلاف‌های حمایتی برای فناوری نوظهور دیده شده‌است. اما در بسیاری موارد این موضوع به تنهایی کافی نیست و همان‌طور که میلن و فارلا (۲۰۱۳) و کیویما و کرن (۲۰۱۶) اشاره نموده‌اند، ضروری است تا از رژیم حاکم مشروعیت‌زدایی شده و حمایت‌ها از آن کاسته شود [۵۲ و ۵۳]. همچنین با قرار دادن فشار بر روی رژیم حاکم از طریق مالیات‌ها، محدودیت واردات یا محدودیت استفاده از فناوری، آنها را به تغییر وادار نمود.

بهره‌برداری از رژیم حاکم (F10):

هر چند در بیشتر موارد رژیم حاکم تمایل به تغییر ندارد اما با این وجود می‌توان ظرفیت‌های بسیار مناسبی در رژیم حاکم برای توسعه‌ی فناوری نوظهور یافت. ماکیتیه و همکاران (۲۰۱۸) شرایطی که در ذیل آن یک بخش می‌تواند تاثیرات مثبتی بر توسعه‌ی نظام نوآوری فناورانه بگذارد را مطرح نموده و برخی کارکردها و سیاست‌های حمایتی برای این منظور را بیان نموده‌اند [۴۹]. بنابراین در کنار تضعیف رژیم موجود، باید ظرفیت‌های آن را برای توسعه‌ی فناوری جدید اهرم نمود و از آن استفاده نمود. مواردی مانند درگیر کردن رژیم

۷

حاکم در توسعه‌ی فناوری جدید، استفاده از زیرساخت‌ها و منابع آن در این امر و بهره‌گیری از فناوری جدید در رفع مشکلات رژیم حاکم می‌تواند به توسعه‌ی نظام نوآوری فناورانه سرعت بخشد.

## ۳ روش تحقیق

- **روش پویایی‌های سیستم**

پویایی‌های سیستم توسط جی فارستر در دانشکده مدیریت سولان دانشگاه ام آی تی[1] طی دهه ۱۹۵۰ توسعه پیدا کرد تا به حل مسایل مدیریتی کمک کند [۵۴]. پویایی‌های سیستم یک روش‌شناسی برای شبیه‌سازی، تحلیل و ارتقای سیستم‌های پویای اجتماعی، اقتصادی و مدیریتی، با استفاده از یک دیدگاه بازخوردی است [۵۵]. پویایی‌های سیستم روش‌های مناسبی را برای شناسایی الگوی رفتاری سیستم در پرتو تفکر سیستمی، معرفی می‌نماید و به علاوه کاربردهای متنوعی در تحلیل سیستم‌ها به خصوص سیستم‌های اقتصادی-اجتماعی (اقتصاد کلان، خرد و تورم)، فنی مهندسی (مکانیک سیالات، برق و ...)، سیاسی (مدل روابط بین الملل، مدل تاثیرات جنگ و صلح)، فرهنگی و اجتماعی (مدل‌های توسعه‌ی شهری، سیاست‌گذاری علوم و ...) را ارایه می‌دهد [۵۶]. مدوز[2] و همکاران (۱۹۷۴)، طبقه‌بندی‌ای از مدل‌های خروجی[3] در پویای‌های سیستم ارایه کرده‌اند که عبارتند از:

- مطلق با پیش‌بینی‌های دقیق
- شرطی با پیش‌بینی‌های دقیق
- شرطی با پیش‌بینی‌های غیردقیق [۵۷].

از آنجایی که پویایی‌های سیستم در وهله اول روشی برای مدل‌سازی سیاست و کسب و کار از طریق شبیه‌سازی است [۵۸]، تمرکز اصلی آن بر دسته سوم این مدل‌هاست:

مدل‌های شبیه‌سازی‌ای که پیش‌نمایی‌هایی مشروط و غیردقیق از رفتار پویا را فراهم می‌آورند. این امر به این دلیل است که نظام‌های اجتماعی و تجاری بر حسب ماهیت خود، به صورت مطلق، غیرقابل پیش‌بینی هستند.

- **روش دیماتل فازی**

دیماتل روشی جامع است که از آن برای ایجاد و تجزیه تحلیل یک مدل ساختاری که شامل روابط علی و معلولی میان عوامل پیچیده می‌باشد، استفاده می‌گردد [۵۹]. کاربرد روش دیماتل بسیار گسترده‌است و در بسیاری از حوزه‌ها مورد استفاده قرار گرفته‌است؛ حوزه‌های مختلفی مانند ارزیابی مزیت رقابتی در پارک‌های علم و فناوری [۶۰]، ارزیابی شایستگی‌های اصلی، تجزیه و تحلیل راه‌حل‌ها، برنامه‌ریزی صنعتی، تجزیه و تحلیل مسایل تصمیم‌گیری در سطح جهانی و ... [۶۱]. همچنین این روش قابلیت ترکیب شدن با سایر روش‌ها مانند فرایند تحلیل شبکه‌ای را داراست که در بسیاری از تحقیقات مورد استفاده قرار گرفته‌است [۶۲، ۶۳]. این روش را می‌توان در چهار مرحله زیر خلاصه نمود:

---

[1] MIT Solan School of Management
[2] Meadows
[3] Outputs models

۸

۱- محاسبه ماتریس متوسط: اگر بخواهیم نظرات H خبره را در رابطه با n عامل استخراج نماییم، از هر خبره می‌خواهیم تا نظر خود در رابطه با میزان تاثیری که عامل i بر عامل j می‌گذارد را بیان نماید. مقدار مقایسات زوجی میان هر دو عامل که با $a_{ij}$ نمایش داده می‌شود به صورت یک عدد صحیح از ۰ تا ۴ ( ۰ = بی‌تاثیر تا عدد ۴ = تاثیر بسیار زیاد) خواهد بود. امتیازات داده شده توسط هر خبره یک ماتریس نامنفی n×n را به وجود خواهد آورد که آن را به صورت زیر نمایش می‌دهیم:

$$X^k = [x_{ij}^k], 1 \leq k \leq H \qquad (1)$$

لذا $X^1$، $X^2$، ...، $X^H$ ماتریس‌های پاسخ هر یک از H خبره می‌باشند و هر عنصر ماتریس $X^k$ عدد صحیحی است که به صورت $x_{ij}^k$ نمایش داده می‌شود. مقادیر روی قطر ماتریس پاسخ هر خبره $X^k$ برابر صفر خواهد بود. در آخر می‌توانیم مقدار ماتریس متوسط A را برای نظرات همه‌ی خبرگان با محاسبه میانگین از امتیازات داده شده توسط فرمول زیر محاسبه نماییم:

$$a_{ij} = \frac{1}{H} \sum_{k=1}^{H} x_{ij}^k \qquad (2)$$

همچنین به ماتریس متوسط $A = [a_{ij}]$، ماتریس رابطه مستقیم اولیه نیز گفته می‌شود. A نشان‌دهنده‌ی تاثیر مستقیم اولیه‌ای است که بر سایر عوامل اعمال و یا از آنها دریافت می‌شود.

۲- محاسبه ماتریس رابطه مستقیم اولیه نرمال شده: ماتریس رابطه مستقیم اولیه نرمال شده D از طریق نرمال کردن ماتریس متوسط A به صورت زیر به دست می‌آید:

$$s = \max\left(\max_{1 \leq i \leq n} \sum_{j=1}^{n} a_{ij}, \max_{1 \leq j \leq n} \sum_{i=1}^{n} a_{ij}\right) \qquad (3)$$

$$D = \frac{A}{s} \qquad (4)$$

باید توجه داشت که مجموع هر ردیف از ماتریس A مانند I نشان‌دهنده‌ی تاثیر مستقیم کل فاکتور i است که بر سایر عوامل اعمال می‌شود، بنابراین $\max_{1 \leq i \leq n} \sum_{j=1}^{n} a_{ij}$ نشان‌دهنده‌ی تاثیر مستقیم کل عامل با بیشترین تاثیر مستقیم بر سایر عوامل است. به همین شکل مجموع هر ستون از ماتریس A مانند I نشان‌دهنده‌ی تاثیر مستقیم کل دریافت شده توسط عامل i از سایر عوامل می‌باشد که $\max_{1 \leq j \leq n} \sum_{i=1}^{n} a_{ij}$ نشان‌دهنده بیشترین تاثیر مستقیم کل دریافت شده توسط عامل از سایر عوامل می‌باشد. ماتریس D از طریق تقسیم هر عنصر ماتریس A بر مقدار s به دست می‌آید. هر عنصر $d_{ij}$ ماتریس D میان ۰ تا ۱ می‌باشد.

۳- محاسبه ماتریس رابطه کل: کاهش پیوسته اثر غیرمستقیم مسایل در توان‌های بالاتر ماتریس D مانند $D^2$، $D^3$، ...، $D^\infty$ همگرایی جواب‌ها برای ماتریس معکوس را تضمین می‌نماید. باید توجه داشت که:

$$\lim_{m \to \infty} D^m = [0]_{n \times n} \qquad (5)$$

$$\lim_{m \to \infty}(I + D + D^2 + D^3 + ... + D^m) = (I - D)^{-1} \qquad (6)$$

۹

همچنین «0» یک ماتریس تهی n×n و I یک ماتریس واحد n×n می‌باشد. ماتریس رابطه کل T یک ماتریس n×n خواهد بود که به صورت زیر تعریف می‌شود:

$$T = [t_{ij}], i, j = 1, 2, ..., n \quad (7)$$

$$T = D + D^2 + ... + D^m = D(I + D + D^2 + ... + D^{m-1}) = D(I-D)^{-1}, m \to \infty$$

r و c را نیز به صورت یک ماتریس $n \times 1$ تعریف می‌نماییم که نمایانگر مجموع ردیف‌ها و ستون‌های ماتریس رابطه کل خواهد بود:

$$r = [r_i]_{n \times 1} = \left(\sum_{j=1}^{n} t_{ij}\right)_{n \times 1} \quad (8)$$

$$c = [c_j]'_{1 \times n} = \left(\sum_{i=1}^{n} t_{ij}\right)'_{1 \times n} \quad (9)$$

$r_i$ برابر مجموع iامین ردیف از ماتریس رابطه کل T می‌باشد. بنابراین $r_i$ نشان‌دهنده‌ی تاثیر کل عامل i می‌باشد که بر سایر عوامل اعمال شده‌است. این تاثیر شامل تاثیر مستقیم و غیرمستقیم می‌باشد. $c_j$ برابر مجموع jامین ستون از ماتریس رابطه کل T می‌باشد. بنابراین $c_j$ نشان‌دهنده‌ی تاثیر کلی می‌باشد که عامل j از سایر عوامل دریافت کرده‌است. این تاثیر شامل تاثیر مستقیم و غیرمستقیم می‌باشد. بنابراین هنگامی که i = j باشد آنگاه $(r_i+c_i)$ برابر تاثیر کل اعمال شده و دریافت شده توسط عامل i می‌باشد. به بیان دیگر $(r_i+c_i)$ نشان‌دهنده‌ی درجه‌ی اهمیت عامل i در سیستم می‌باشد. همچنین $(r_i-c_i)$ نشان‌دهنده‌ی تاثیر خالصی است که عامل i در کل سیستم اعمال می‌کند. هنگامی که $(r_i-c_i)$ مقداری مثبت باشد به معنای آن است که عامل i در کل یک عامل تاثیرگذار در سیستم می‌باشد و هنگامی که $(r_i-c_i)$ مقداری منفی باشد به این معناست که عامل i در کل یک عامل تاثیرپذیر در سیستم می‌باشد [64].

4- تعیین مقدار آستانه‌ای و به دست آوردن نقشه تاثیر-رابطه: در بسیاری از تحقیقات به منظور نشان دادن رابطه ساختاری میان عوامل، در عین حفظ پیچیدگی سیستم در حد قابل کنترل، نیاز است که مقدار آستانه‌ای p را به گونه‌ای تعیین نماییم که تنها تاثیرات قابل چشم پوشی در ماتریس T را فیلتر نماید. تنها تاثیراتی در ماتریس T که بزرگ‌تر از مقدار آستانه‌ای می‌باشند بایستی انتخاب شده و در نمودار نقشه تاثیر- روابط (IRM) یا روابط علی نمایش داده شوند. معمولا این مقدار آستانه‌ای توسط خبرگان تعیین می‌شود [65].

• **منطق فازی**

مجموعه‌های فازی برای اولین بار توسط زاده در سال 1965 معرفی شد. این نظریه یک ابزار ریاضیاتی جدید برای کار با عدم قطعیت اطلاعات فراهم نمود. از آن زمان تا کنون این نظریه به خوبی توانسته‌است توسعه یابد و کاربردهای موفق واقعی بسیاری پیدا کند. در تعامل با ابهام افکار و بیان انسانی، نظریه مجموعه فازی بسیار راهگشا می‌باشد. به خصوص، هنگام کار با نارسایی‌های موجود در فرایند تخمین‌های زبانی، تبدیل عبارات زبانی به اعداد فازی بسیار مفید می‌باشد. یک متغیر زبانی، متغیری است که مقدار آن دارای یک شکل یا عبارت یا جمله در زبان

طبیعی می‌باشد [۶۶]. در عمل، مقادیر زبانی می‌توانند با استفاده از اعداد فازی نشـان داده شـوند، کـه اعـداد فـازی مثلثی مرسوم‌ترین آنها می‌باشند.

### • عدد فازی مثلثی

اعداد فازی زیر مجموعه‌ای از اعداد حقیقی هستند که در واقع از ایـده فاصـله اطمینـان بسـط یافتـه‌انـد. بـر اسـاس تعریف، عدد فازی $A$ روی $R$ یک عدد فـازی مثلثی (TFN) است هرگـاه تـابع عضـویت $\mu_{\tilde{A}}(x): R \to [0,1]$، به‌صورت زیر باشد که $L$ و $U$ به ترتیب حد پایین و بالای عدد فازی $\tilde{A}$ می‌باشند [۶۷].

$$\mu_{\tilde{A}}(x) = \begin{cases} (x-L)/(M-L), & L \leq x \leq M \\ (U-x)/(U-M) & M \leq x \leq U \\ 0 & otherwise \end{cases} \quad (10)$$

### • کاربرد منطق فازی در روش دیماتل

یکی از مسایل در استفاده از روش دیماتل به دست آوردن اندازه تاثیر مستقیم میان هر دو عامل می‌باشد. انـدازه‌ی این امتیازات همواره با استفاده از پیمایش خبرگان به دست مـی‌آیـد؛ امـا در بسـیاری از مـوارد قضـاوت افـراد در تصمیم‌گیری غیرواضح می‌باشد و اندازه‌گیری آنها بـا اسـتفاده از مقـادیر عـددی دقیـق میسـر نمـی‌باشـد؛ بنـابراین استفاده از منطق فازی در کار کردن با مسایلی که از مشخصه‌هایشان ابهام و عدم دقت می باشد، ضروری می‌باشد. از این رو، نیاز به توسعه روش دیماتل با استفاده از منطق فازی به منظور تصمیم‌گیری بهتر در محیط فازی احسـاس می‌شود.

به منظور استفاده از منطق فازی در روش دیماتل باید در گام نخست این روش که در آن نظرات خبرگان در رابطه با میزان تاثیر عوامل بر یکدیگر احصا می‌شوند از خبرگان درخواست نمود تا بـر اسـاس متغیرهـای زبـانی تعریـف شده پاسخ دهند. برای غلبه کردن بر ابهامات ارزیابی‌های انسانی، از متغیر زبانی «تاثیر» بـا اسـتفاده از پـنج عبـارت: خیلی‌زیاد، زیاد، کم، خیلی‌کم و بی‌تاثیر که به صورت اعداد فازی مثلثی مثبت $(l_{ij}, m_{ij}, r_{ij})$ بیان می‌شـوند [۶۸] – همان‌گونه که در جدول ۱ نشان داده شده‌است - استفاده می‌شـود. پـس از اخـذ نظـرات خبرگـان، بـا اسـتفاده از میانگین‌گیری فازی، ماتریس متوسط فازی را محاسبه می‌نماییم. سپس بـا اسـتفاده از روابـط موجـود بـرای تبـدیل مقادیر فازی به اعداد غیرفازی ماتریس مقادیر متوسط نهایی به دست می‌آید. ادامه رونـد هماننـد آنچـه کـه پیشـتر توضیح داده شد می‌باشد [۶۹].

**جدول ۱.** مقیاس زبانی فازی

| عبارت‌های زبانی | اعداد فازی مثلثی |
|---|---|
| تاثیر خیلی‌زیاد | (۰/۷۵، ۱، ۱) |
| تاثیر زیاد | (۰/۵، ۰/۷۵، ۱) |
| تاثیر کم | (۰/۲۵، ۰/۵، ۰/۷۵) |
| تاثیر خیلی‌کم | (۰، ۰/۲۵، ۰/۵) |
| بی‌تاثیر | (۰، ۰، ۰/۲۵) |

بنابراین اگر $p$ پاسخ‌دهنده داشته باشیم به تعداد پاسخ‌دهندگان، ماتریس‌های فازی $\tilde{z}^1, \tilde{z}^2, ..., \tilde{z}^p$ خواهیم داشـت. در

۱۱

نتیجه ماتریس میانگین فازی با استفاده از رابطه (11) محاسبه می‌شود:

$$\tilde{z} = (\tilde{z}^1 + \tilde{z}^2 + ... + \tilde{z}^p)/p \quad (11)$$

به منظور دیفازی کردن مقادیر، از روش $CFCS^1$ استفاده می‌شود. روش CFCS براساس محاسبه مقادیر راست و چپ با استفاده از مینیمم و ماکسیمم فازی می‌باشد. در این روش مقدار نهایی به عنوان یک متوسط وزنی مطابق با تابع عضویت تعریف می‌شود [70]. اگر $(l_{ij}, m_{ij}, r_{ij})$ نشان‌دهنده‌ی میزان تاثیر معیار i بر روی معیار j در ماتریس فازی رابطه مستقیم اولیه باشد؛ آنگاه روش CFCS را می‌توان در مراحل زیر خلاصه نمود:

گام نخست نرمال‌سازی می‌باشد:

$$xl_{ij} = (l_{ij} - \min l_{ij}) / \Delta_{\min}^{\max}$$
$$xm_{ij} = (m_{ij} - \min l_{ij}) / \Delta_{\min}^{\max} \qquad \Delta_{\min}^{\max} = \max r_{ij} - \min l_{ij} \quad (12)$$
$$xr_{ij} = (r_{ij} - \min l_{ij}) / \Delta_{\min}^{\max}$$

در گام دوم به محاسبه‌ی مقادیر سمت راست و سمت چپ می‌پردازیم:

$$xls_{ij} = xm_{ij} / (1 + xm_{ij} - xl_{ij})$$
$$xrs_{ij} = xr_{ij} / (1 + xr_{ij} - xm_{ij}) \quad (13)$$

در گام سوم به محاسبه‌ی مقدار قطعی نرمال شده‌ی کل می‌پردازیم:

$$x_{ij} = [xls_{ij}(1 - xls_{ij}) + xrs_{ij} xrs_{ij}] / [1 - xls_{ij} + xrs_{ij}] \quad (14)$$

و در گام آخر به محاسبه‌ی مقدار قطعی نهایی می‌پردازیم:

$$z_{ij} = \min l_{ij} + x_{ij} \Delta_{\min}^{\max} \quad (15)$$

به منظور انجام محاسبات ذکر شده از نرم افزار متلب استفاده گردید.

### • پیمایش خبرگان

در این تحقیق از پیمایش خبرگان برای شناسایی اثرات عوامل دهگانه بر روی یکدیگر استفاده می‌شود. خبرگان مورد استفاده در این تحقیق باید دارای سه ویژگی آشنایی با علم و فناوری اینترنت اشیا، شناخت مناسب شرکت‌های دانش بنیان و صنعتی فعال در حوزه‌ی اینترنت اشیا و تجربه‌ی سیاستگذاری در نهادهای عمومی کشور باشند. با توجه به این ویژگی‌ها جامعه‌ی خبرگان این تحقیق در کشور از پیش مشخص شده نیستند. در تحقیقاتی که دسترسی به نمونه آماری دارای ویژگی‌های مورد نظر، دشوار است و یا اینکه کمیاب باشند از روش نمونه‌گیری گلوله‌ی برفی استفاده می‌شود [71،72]. لذا به منظور شناسایی خبرگان این حوزه از روش گلوله برفی استفاده گردید. تعداد خبرگان شناسایی‌شده چهارده نفر می‌باشند که تحصیلات دانشگاهی مرتبط با رشته‌های الکترونیک، مخابرات و فناوری اطلاعات را دارا می‌باشند و سابقه‌ی همکاری با نهادهای سیاستگذار حاکمیتی مانند معاونت علمی و فناوری ریاست جمهوری، مرکز همکاری‌های تحول و پیشرفت، شورای عالی فضای مجازی، پژوهشگاه نیرو و پژوهشگاه ارتباطات و فناوری اطلاعات و ... را دارا می‌باشند.

---

[1] Converting Fuzzy data into Crisp Scores

**٤ نتایج**

به منظور ارایه‌ی مدل پویای تعاملات کارکردی نظام نوآوری فناورانه‌ی اینترنت اشیا در گام نخست با استفاده از پیمایش خبرگان اثرات عوامل دهگانه بر روی یکدیگر استخراج گردید. برای این منظور از آنها خواسته شد تا با استفاده از متغیرهای زبانی میزان تاثیر هر یک از عوامل ۱۰ گانه بر روی یکدیگر را بیان نمایند. به منظور طراحی مدل پویای تعاملات کارکردی با استفاده از روش پویایی‌های سیستم با معادل‌سازی متغیرهای زبانی با اعداد صحیح - صفر به معنی بی‌تاثیر تا ۴ به معنای تاثیر خیلی‌زیاد- و میانگین‌گیری نظرات خبرگان، اثرگذاری عوامل بر یکدیگر شناسایی شدند. با در نظر گرفتن میانگین بیش از ۳ به معنی اثرگذاری زیاد، بیان می‌داریم یک عامل بر عامل دیگر اثر می‌گذارد. در غیر این صورت بین دو عامل اثری در نظر گرفته نمی‌شود.

مدل سیستم داینامیک در شکل ۱ قابل مشاهده می‌باشد. همان‌طور که می‌بینیم، عامل شکل‌گیری بازار با تاثیرگذاری مستقیم بر ۶ عامل دیگر، بیش‌ترین تعداد تاثیرگذاری مستقیم را بر سایر عوامل دارد و پس از آن عامل بسیج منابع و بهره‌برداری از رژیم موجود با تاثیرگذاری بر ۵ عامل، قرار دارند. باید توجه داشت که این تاثیرگذاری مستقیم می‌باشد و در یک سیستم ممکن است تاثیر نهایی کل شامل تاثیر مستقیم و غیرمستقیم متفاوت باشد. برای همین منظور از روش دیماتل فازی برای یافتن تاثیر نهایی عوامل بر روی یکدیگر استفاده می‌نماییم.

**شکل ۱.** مدل پویایی‌های سیستم نظام نوآوری فناورانه اینترنت اشیا

برای این منظور با استفاده از تبدیل عبارت‌های زبانی به اعداد فازی مثلثی، ابتدا ماتریس تاثیر اولیه فازی را ایجاد می‌نماییم. سپس با استفاده از روش CFCS و بر اساس روابط (۱۲) تا (۱۵)، ماتریس اولیه، غیرفازی می‌شود که این ماتریس با دقت دو رقم اعشار در جدول ۲ قابل مشاهده می‌باشد.

**جدول ۲.** ماتریس اولیه غیرفازی

| | F1 | F2 | F3 | F4 | F5 | F6 | F7 | F8 | F9 | F10 | SUM |
|---|---|---|---|---|---|---|---|---|---|---|---|
| **F1** | ۰/۰۰ | ۰/۵۳ | ۰/۷۹ | ۰/۴۷ | ۰/۶۴ | ۰/۶۵ | ۰/۳۲ | ۰/۶۲ | ۰/۴۶ | ۰/۶۸ | ۵/۱۶ |
| **F2** | ۰/۶۷ | ۰/۰۰ | ۰/۵۲ | ۰/۸۷ | ۰/۵۰ | ۰/۸۴ | ۰/۵۳ | ۰/۴۴ | ۰/۵۵ | ۰/۷۳ | ۵/۶۵ |
| **F3** | ۰/۸۹ | ۰/۶۲ | ۰/۰۰ | ۰/۵۹ | ۰/۷۰ | ۰/۶۵ | ۰/۵۸ | ۰/۵۳ | ۰/۶۰ | ۰/۵۵ | ۵/۷۱ |
| **F4** | ۰/۵۸ | ۰/۶۹ | ۰/۷۸ | ۰/۰۰ | ۰/۷۵ | ۰/۸۷ | ۰/۶۵ | ۰/۸۰ | ۰/۷۱ | ۰/۷۳ | ۶/۵۶ |
| **F5** | ۰/۹۱ | ۰/۶۹ | ۰/۵۸ | ۰/۷۸ | ۰/۰۰ | ۰/۶۱ | ۰/۵۰ | ۰/۶۱ | ۰/۸۱ | ۰/۸۷ | ۶/۳۶ |
| **F6** | ۰/۶۴ | ۰/۶۴ | ۰/۴۷ | ۰/۶۱ | ۰/۷۲ | ۰/۰۰ | ۰/۷۸ | ۰/۵۹ | ۰/۷۰ | ۰/۵۸ | ۵/۷۳ |
| **F7** | ۰/۵۳ | ۰/۴۷ | ۰/۶۴ | ۰/۶۷ | ۰/۷۲ | ۰/۷۱ | ۰/۰۰ | ۰/۸۷ | ۰/۵۳ | ۰/۷۴ | ۵/۸۸ |
| **F8** | ۰/۷۶ | ۰/۶۶ | ۰/۵۹ | ۰/۵۰ | ۰/۶۰ | ۰/۶۱ | ۰/۵۹ | ۰/۰۰ | ۰/۵۴ | ۰/۶۳ | ۵/۴۸ |
| **F9** | ۰/۴۹ | ۰/۴۰ | ۰/۶۵ | ۰/۶۴ | ۰/۷۶ | ۰/۵۳ | ۰/۳۵ | ۰/۳۴ | ۰/۰۰ | ۰/۵۹ | ۴/۷۵ |
| **F10** | ۰/۸۰ | ۰/۷۳ | ۰/۸۴ | ۰/۷۳ | ۰/۸۷ | ۰/۷۱ | ۰/۴۶ | ۰/۵۴ | ۰/۲۸ | ۰/۰۰ | ۵/۹۶ |
| **SUM** | ۶/۲۷ | ۵/۴۳ | ۵/۸۶ | ۵/۸۶ | ۶/۲۶ | ۶/۱۸ | ۴/۷۶ | ۵/۳۴ | ۵/۱۸ | ۶/۱ | |

طبق رابطه‌ی (۳) مقدار ۶/۵۶= S خواهد بود. با استفاده از رابطه‌ی (۴) ماتریس تاثیر مستقیم اولیه نرمال شده محاسبه می‌گردد. این ماتریس در جدول ۳ قابل مشاهده می‌باشد:

**جدول ۳.** ماتریس تاثیر مستقیم اولیه نرمال شده

| | F1 | F2 | F3 | F4 | F5 | F6 | F7 | F8 | F9 | F10 |
|---|---|---|---|---|---|---|---|---|---|---|
| **F1** | ۰/۰۰ | ۰/۰۸ | ۰/۱۲ | ۰/۰۷ | ۰/۱۰ | ۰/۱۰ | ۰/۰۵ | ۰/۹ | ۰/۰۷ | ۰/۱۰ |
| **F2** | ۰/۱۰ | ۰/۰۰ | ۰/۰۸ | ۰/۱۳ | ۰/۰۸ | ۰/۱۳ | ۰/۰۸ | ۰/۷ | ۰/۰۸ | ۰/۱۱ |
| **F3** | ۰/۱۴ | ۰/۹ | ۰/۰۰ | ۰/۹ | ۰/۱۱ | ۰/۱۰ | ۰/۹ | ۰/۸ | ۰/۹ | ۰/۰۸ |
| **F4** | ۰/۹ | ۰/۱۱ | ۰/۱۲ | ۰/۰۰ | ۰/۱۱ | ۰/۱۳ | ۰/۱۰ | ۰/۱۲ | ۰/۱۱ | ۰/۱۱ |
| **F5** | ۰/۱۴ | ۰/۱۱ | ۰/۰۹ | ۰/۱۲ | ۰/۰۰ | ۰/۰۹ | ۰/۰۸ | ۰/۹ | ۰/۱۲ | ۰/۱۳ |
| **F6** | ۰/۱۰ | ۰/۱۰ | ۰/۰۷ | ۰/۰۹ | ۰/۱۱ | ۰/۰۰ | ۰/۱۲ | ۰/۹ | ۰/۱۱ | ۰/۰۹ |
| **F7** | ۰/۰۸ | ۰/۰۷ | ۰/۱۰ | ۰/۱۰ | ۰/۱۱ | ۰/۱۱ | ۰/۰۰ | ۰/۱۳ | ۰/۰۸ | ۰/۱۱ |
| **F8** | ۰/۱۲ | ۰/۱۰ | ۰/۰۹ | ۰/۰۸ | ۰/۰۹ | ۰/۰۹ | ۰/۰۹ | ۰/۰۰ | ۰/۰۸ | ۰/۱۰ |
| **F9** | ۰/۰۸ | ۰/۰۶ | ۰/۱۰ | ۰/۱۰ | ۰/۱۲ | ۰/۰۸ | ۰/۰۵ | ۰/۰۵ | ۰/۰۰ | ۰/۰۹ |
| **F10** | ۰/۱۲ | ۰/۱۱ | ۰/۱۳ | ۰/۱۱ | ۰/۱۳ | ۰/۱۱ | ۰/۰۷ | ۰/۰۸ | ۰/۰۴ | ۰/۰۰ |

سپس با استفاده از رابطه‌ی (۸) ماتریس رابطه‌ی کل محاسبه شد.

۱۴

**جدول ٤.** ماتریس رابطه‌ی کل

|     | F1   | F2   | F3   | F4   | F5   | F6   | F7   | F8   | F9   | F10  |
|-----|------|------|------|------|------|------|------|------|------|------|
| F1  | ۰/۶۱ | ۰/۶۱ | ۰/۶۷ | ۰/۶۳ | ۰/۶۹ | ۰/۶۹ | ۰/۵۲ | ۰/۶۱ | ۰/۵۷ | ۰/۶۸ |
| F2  | ۰/۷۵ | ۰/۵۸ | ۰/۶۹ | ۰/۷۴ | ۰/۷۳ | ۰/۷۷ | ۰/۵۹ | ۰/۶۳ | ۰/۶۳ | ۰/۷۴ |
| F3  | ۰/۷۸ | ۰/۶۷ | ۰/۶۲ | ۰/۷۰ | ۰/۷۵ | ۰/۷۴ | ۰/۵۹ | ۰/۶۴ | ۰/۶۴ | ۰/۷۲ |
| F4  | ۰/۸۴ | ۰/۷۶ | ۰/۸۱ | ۰/۷۱ | ۰/۸۵ | ۰/۸۶ | ۰/۶۸ | ۰/۷۶ | ۰/۷۳ | ۰/۸۳ |
| F5  | ۰/۸۵ | ۰/۷۴ | ۰/۷۷ | ۰/۷۹ | ۰/۷۳ | ۰/۸۰ | ۰/۶۳ | ۰/۷۱ | ۰/۷۲ | ۰/۸۳ |
| F6  | ۰/۷۶ | ۰/۶۷ | ۰/۶۹ | ۰/۷۱ | ۰/۷۶ | ۰/۶۶ | ۰/۶۲ | ۰/۶۶ | ۰/۶۶ | ۰/۷۳ |
| F7  | ۰/۷۶ | ۰/۶۷ | ۰/۷۳ | ۰/۷۳ | ۰/۷۸ | ۰/۷۷ | ۰/۵۳ | ۰/۷۱ | ۰/۶۵ | ۰/۷۷ |
| F8  | ۰/۷۴ | ۰/۶۵ | ۰/۶۸ | ۰/۶۷ | ۰/۷۲ | ۰/۷۱ | ۰/۵۸ | ۰/۵۵ | ۰/۶۱ | ۰/۷۱ |
| F9  | ۰/۶۴ | ۰/۵۶ | ۰/۶۲ | ۰/۶۲ | ۰/۶۷ | ۰/۶۳ | ۰/۴۹ | ۰/۵۴ | ۰/۴۸ | ۰/۶۳ |
| F10 | ۰/۸۱ | ۰/۷۲ | ۰/۷۷ | ۰/۷۵ | ۰/۸۱ | ۰/۷۹ | ۰/۶۱ | ۰/۶۸ | ۰/۶۳ | ۰/۶۸ |

با استفاده از ماتریس رابطه کل و با استفاده از روابط (۸) و (۹) مقادیر $r_i$ و $c_j$ محاسبه شـدند. مقـادیر $r_i$، $c_j$، $(r_i + c_j)$ و $(r_i - c_j)$ در جدول ۵ آورده شده‌اند:

**جدول ۵.** مقادیر $r_i$، $c_j$، $(r_i+ c_j)$ و $(r_i- c_j)$

|     | ri    | cj    | (ri+ cj) | (ri- cj) |
|-----|-------|-------|----------|----------|
| F1  | ۶/۲۸  | ۷/۵۵  | ۱۳/۸۳    | -۱/۲۷    |
| F2  | ۶/۸۵  | ۶/۶۳  | ۱۳/۴۸    | ۰/۲۲     |
| F3  | ۶/۸۷  | ۷/۰۵  | ۱۳/۹۲    | -۰/۱۸    |
| F4  | ۷/۸۳  | ۷/۰۴  | ۱۴/۸۷    | ۰/۷۹     |
| F5  | ۷/۵۷  | ۷/۴۹  | ۱۵/۰۶    | ۰/۰۸     |
| F6  | ۶/۹۲  | ۷/۴۳  | ۱۴/۳۴    | -۰/۵۱    |
| F7  | ۷/۱۱  | ۵/۸۳  | ۱۲/۹۵    | ۱/۲۸     |
| F8  | ۶/۶۱  | ۶/۴۸  | ۱۳/۰۸    | ۰/۱۳     |
| F9  | ۵/۸۷  | ۶/۳۲  | ۱۲/۱۹    | -۰/۴۶    |
| F10 | ۷/۲۴  | ۷/۳۳  | ۱۴/۵۷    | -۰/۰۹    |

بنابراین همان‌طور که در جدول ۵ نیز مشاهده می‌شود، عوامل فعالیت‌های کارآفرینانه، شکل‌گیـری بـازار، بسیج منابع، سیاستگذاری و ایجاد هماهنگی و ایجاد ساختار به صورت کلی در سیستم اثرگذار می‌باشند و سـایر عوامـل شامل توسعه، تبادل و انتشار دانش، جهت‌دهی به تحقیقات، مشروعیت‌بخشی، تضعیف رژیم موجود و بهره‌بـرداری از رژیم موجود تاثیرپذیر می‌باشند. بنابراین سیاستگذار باید توجه ویژه‌ی خود را بـه عوامـل تاثیرگـذار در سیسـتم معطوف بدارد و از این طریق به عملکرد بهتر سایر عوامل و نیز عملکرد کلی سیستم دست یابد.

## ۵ جمع‌بندی

فناوری اینترنت اشیا یکی از مهم‌ترین پیشران‌هـای توسـعه در آینـده‌ی نزدیـک خواهـد بـود و تـاثیر شـگرفی بـر توسعه‌ی اقتصادی کشورها ایجاد خواهد نمود. به همین منظور سیاستگذاری این فناوری در دستور کـار دولت‌هـا

قرار گرفته‌است. یکی از مهم‌ترین الگوهای سیاستگذاری برای فناوری‌هـای نـوین، روش نظـام نـوآوری فناورانـه می‌باشد. این روش با سابقه‌ای طولانی در سیاستگذاری فناوری‌های پیشرفته توانسته‌است خطـوط راهنمـای مفیـدی برای توسعه‌ی فناوری‌های نوین مطرح نماید. در همین راستا بر اساس ادبیات موضوع در این حوزه عوامل اصـلی تاثیرگذار بر توسعه‌ی فناوری‌های نوظهور شناسایی شـدند. سـپس بـا اسـتفاده از روش پویایی‌هـای سیسـتم، مـدل تعاملات میان این عوامل شناسایی گردید. در این مدل 20 حلقه‌ی بازخوردی شناسایی گردید. سپس با اسـتفاده از روش دیماتل فازی میزان اثرگذاری عوامل شناسایی شدند. خلاصه‌ی نتایج روش پویایی سیستم‌ها و دیماتل فازی در جدول 6 قابل مشـاهده مـی‌باشـد. در ایـن جـدول براسـاس میـزان اثرگـذاری و تعـداد حلقـه هـای بـازخوردی کارکردها رتبه‌بندی شده‌اند.

**جدول 6.** نحوه‌ی اثرگذاری، میزان تعاملات و تعداد حلقه‌های بازخوردی هر یک از کارکردها

| | تعداد روابط | تعداد حلقه‌های بازخوردی | نحوه اثرگذاری در سیستم |
|---|---|---|---|
| شکل گیری بازار | 6 | 11 | تاثیرگذار |
| بسیج منابع | 5 | 11 | تاثیرگذار |
| فعالیت های کارآفرینانه | 3 | 9 | تاثیرگذار |
| سیاستگذاری و ایجاد هماهنگی | 2 | 5 | تاثیرگذار |
| ایجاد ساختار | 1 | 0 | تاثیرگذار |
| بهره برداری از رژیم موجود | 5 | 14 | تاثیرپذیر |
| مشروعیت بخشی | 2 | 6 | تاثیرپذیر |
| جهت دهی به تحقیقات | 1 | 1 | تاثیرپذیر |
| تضعیف رژیم موجود | 1 | 1 | تاثیرپذیر |
| توسعه، تبادل و انتشار دانش | 1 | 1 | تاثیرپذیر |

در میان عوامل، چهار عامل شکل‌گیری بازار، بسیج منابع، بهـره‌بـرداری از رژیـم موجـود و سیاسـتگذاری و ایجاد هماهنگی بیش‌ترین تـاثیر مسـتقیم را بـر سـایر عوامـل دارا هسـتند. همچنـین 5 عامـل سیاسـتگذاری و ایجـاد هماهنگی، شکل‌گیری بازار، فعالیت‌های کارآفرینانه، ایجاد سـاختار و بسیج منـابع اثرگـذار مـی‌باشـند. در میـان عوامل تاثیرگذار بسیج منابع، ایجـاد سـاختار و فعالیـت‌هـای کارآفرینانـه از تـاثیر محـدودی بـر روی کـل سیسـتم برخوردار هستند. در میان عوامل تاثیرگذار ایجاد ساختار از تاثیر محدودی بر روی کل سیسـتم برخـوردار اسـت و تعداد روابط و حلقه های بازخوردی آن بسیار کم می‌باشد. بنابراین با توجه به محـدودیت منـابع در یـک سیسـتم، پرداختن به این عامل آخرین اولویت را در میان عوامل تاثیرگذار دارا می‌باشد. بنابراین عمده‌ی توجـه سیاسـتگذار باید بر چهار عامل دیگر باشد. عامل شکل‌گیری بازار هـم در تـاثیر مستقیم یکـی از بـیش‌تـرین میـزان تـاثیرات را برخوردار می‌باشد و هم در تاثیرکل که دربرگیرنده‌ی تاثیر مستقیم و غیرمستقیم بر سایر عوامـل می‌باشـد جایگـاه بالایی دارد. همچنین از منظر روابط و حلقـه هـای بـازخوردی نیـز بـالاترین جایگـاه را دارا مـی باشـد. بنـابراین، فعالیت‌هایی مانند خریدهای عمومی، مشوق‌هـای مالیـاتی، جـوایز صـادراتی، نظـارت بـر سـهم واردات، توسـعه‌ی اندازه‌ی بازار و سرمایه‌گذاری در کسب و کارهای جدید کـه در ذیـل شـکل گیـری بـازار تعریـف مـی‌شـوند از

بیش‌ترین اهمیت برخوردار خواهند بود. در بسیج منابع توجه تامین انواع وام و سرمایه‌های خطرپذیر، ایجاد زیرساخت‌های مختلف فنی و آزمایشگاهی نقش مهمی در توسعه‌ی فناوری خواهند داشت. در فعالیت‌های کارآفرینانه مواردی همچون انتقال فناوری و پیمان‌های استراتژیک با شرکت‌های بین‌المللی، تمرکز بر صادرات و ایجاد تنوع در کاربردها و حوزه‌های کاری شرکت‌ها توانایی ایجاد تحول در این حوزه را دارا می‌باشند. همچنین، پرداختن به مواردی مانند ایجاد مرکز فرماندهی، سیاست‌گذاری مشارکتی، تدوین و اصلاح مقررات، ایجاد هم‌راستایی در قوانین در سطوح مختلف، ایجاد هماهنگی میان بازیگران مختلف، هم‌راستاسازی علایق و منافع و رتبه‌بندی شرکت‌ها و محصولات آنها نیز که در ذیل کارکرد سیاست‌گذاری و ایجاد هماهنگی تعریف می‌شوند، در گام بعد می‌تواند سبب سرعت بخشیدن به توسعه‌ی فناوری اینترنت اشیا در کشور شود. این کارکرد بیش‌ترین تاثیرگذاری را در کل سیستم را دارا می‌باشد اما چون از منظر تعداد روابط و حلقه‌های بازخوردی در جایگاه پایین‌تری نسبت به سایر کارکردها قرار دارد، به صورت کلی در جایگاه پایین‌تری از منظر سیاست‌گذاری قرار می‌گیرد. همان‌طور که مشخص است عمده‌ی فعالیت‌هایی که در ذیل این کارکردها مطرح شده‌اند کارهای حاکمیتی می‌باشند که بدون تخصیص بودجه‌های جدید امکان پیگیری دارند و عمدتا از طریق هدایت بودجه‌های موجود و یا ایجاد هماهنگی در بخش‌های مختلف قابل پیگیری می‌باشند. لذا نهادسازی به منظور پیگیری این امور، می‌تواند بسیار راهگشا بوده و منجر به توسعه‌ی این فناوری در کشور شود. یکی از پویاترین نهادهایی که در سال‌های گذشته در کشور به منظور توسعه‌ی فناوری‌های نوین شکل گرفته‌است، ستادهای توسعه‌ی فناوری نوین بوده‌اند که شکل‌گیری یک ستاد برای توسعه‌ی فناوری اینترنت اشیا نیز می‌تواند گامی موثر در این راستا باشد.

باید توجه داشت که هم‌افزایی میان همه‌ی عوامل است که منجر به توسعه‌ی کامل سیستم خواهد شد، اما از منظر سیاست‌گذار که با محدودیت جدی منابع روبه‌رو می‌باشد، انتخاب عواملی که پرداختن به آنها منجر به اثر آبشاری بر روی سایر عوامل شده و حلقه‌های بازخوردی ایجاد نماید که سایر عوامل را نیز ارتقا بخشد از درجه‌ی بالاتری از اهمیت برخوردار است. بنابراین اگر محدودیت منابع کاهش یابد و امکان پرداختن به سایر عوامل نیز برای سیاست‌گذار فراهم باشد، ضروری است تا به تحریک سایر عوامل نیز پرداخته شود. لذا در این تحقیق برخلاف تحقیقات پیشین حوزه که پیشنهادات سیاستی در رابطه با همه‌ی کارکردها مطرح می‌شود، موضوع اولویت‌گذاری به علت وجود محدودیت منابع در نظر گرفته شده است و با تمرکز بر کارکردهای اثرگذار، سعی شده است تا از خاصیت نقطه‌ی اهرمی استفاده شده و با صرف کمترین منابع توسط سیاست‌گذار، بیش‌ترین بازدهی برای سیستم حاصل گردد.

همچنین با در نظر گرفتن چرخه‌ی عمر فناوری (شکل‌گیری، رشد، بلوغ و افول) و چرخه‌ی توسعه‌ی نظام نوآورانه فناوری باید توجه داشت که تکامل ساختاری منجر به ایجاد نقاط قوت و ضعف جدیدی در کارکردها می‌شود و بنابراین در هر مرحله از این چرخه نیازمند ترکیب سیاستی اصلاح شده‌ای می‌باشیم [73، 74]، لذا با گذشت زمان ممکن است پویایی‌های موجود در سیستم تغییر نموده و روابط میان عوامل دگرگون شود. بنابراین رصد دایمی سیستم برای سیاست‌گذار امری ضروری بوده که باید به صورت مستمر صورت پذیرد و از این طریق

در هر دوره‌ی خاص مهم‌ترین عوامل برای توسعه‌ی فناوری اینترنت اشیا متناسب با شرایط روز آن شناسایی گردند.

به منظور تحقیقات آتی در این حوزه موارد مختلفی پیشنهاد می‌گردد. در اولین گام می‌توان از این روش در سیاست‌گذاری سایر فناوری‌های نوین نیز استفاده نمود. همچنین استفاده از سایر روش‌هایی که به شناسایی روابط میان عوامل در یک سیستم می‌پردازند مانند روش معادلات ساختاری و یا استفاده از سایر روش‌های تصمیم‌گیری که به اولویت‌بندی عوامل موثر در یک سیستم می‌پردازند مانند روش فرایند تحلیل شبکه‌ای و مقایسه‌ی نتایج آن با این روش نیز می‌تواند دانش‌افزایی جدیدی در این حوزه داشته باشد. بررسی پویایی‌های درون کارکردی نیز می‌تواند قدم بعدی تحقیقات در این حوزه باشد [۷۵]. از آنجایی که هر کارکرد خود شامل زیرکارکردهای متنوعی است، به دست آوردن تعاملات میان آنها و اولویت‌بندی آنها نیز می‌تواند یک گام رو به جلو به منظور بالا بردن کیفیت و دقت سیاست‌گذاری‌ها باشد.

## منابع

موسی خانی و همکاران، ارایه‌ی مدل پویای تعاملات کارکردی نظام نوآوری فناورانه‌ی اینترنت اشیا با استفاده از پویایی‌های سیستم و دیماتل فازی[31] Salie, F., de Jager, K., Dreher, C., Douglas, T. S., (2019). The scientific base for orthopaedic device development in South Africa: spatial and sectoral evolution of knowledge development. Scientometrics, 119(1), 31-54.

[32] Zainab, A., Liaquat, R., Meraj, S., (2019). Major barriers to the diffusion of bio-digestion technology in Pakistan. International Journal of Energy Sector Management, ahead-of-print(ahead-of-print).

[33] Liu, G., Gao, P., Chen, F., Yu, J., Zhang, Y., (2018). Technological innovation systems and IT industry sustainability in China: A case study of mobile system innovation. Telematics and Informatics, 35(5), 1144-1165.

[34] Gandenberger, C., Strauch, M., (2018). Wind energy technology as opportunity for catching-up? A comparison of the tis in Brazil and China. Innovation and Development, 8(2), 287-308.

[35] Azad, S. M., Ghodsypour, S. H., (2018a). Modeling the dynamics of technological innovation system in the oil and gas sector. Kybernetes, 47(4), 771-800.

[36] Azad, S. M., Ghodsypour, S. H., (2018b). A novel system dynamics' model for the motors of a sectoral innovation system–simulating and policymaking. African Journal of Science, Technology, Innovation and Development, 10(3), 239-257.

[37] Nasiri, M., Ramazani Khorshid-Doust, R., Bagheri Moghaddam, N., (2015). The status of the hydrogen and fuel cell innovation system in Iran. Renewable and Sustainable Energy Reviews, 43, 775-783.

[38] Moallemi, E. A., Ahamdi, A., Afrazeh, A., Bagheri Moghaddam, N., (2014). Understanding systemic analysis in the governance of sustainability transition in renewable energies: The case of fuel cell technology in Iran. Renewable and Sustainable Energy Reviews, 33, 305-315.

[39] Nasiri, M., Ramazani Khorshid-Doust, R., Bagheri Moghaddam, N., (2013). Effects of under-development and oil-dependency of countries on the formation of renewable energy technologies: A comparative study of hydrogen and fuel cell technology development in Iran and the Netherlands. Energy Policy, 63, 588-598.

[40] Wieczorek, A. J., Negro, S. O., Harmsen, R., Heimeriks, G. J., Luo, L., Hekkert, M. P., (2013). A review of the European offshore wind innovation system. Renewable and Sustainable Energy Reviews, 26, 294-306.

[41] Andreasen, K. P., Sovacool, B. K., (2015). Hydrogen technological innovation systems in practice: comparing Danish and American approaches to fuel cell development. Journal of Cleaner Production, 94, 359-368.

[42] Perez Vico, E., (2014). An in-depth study of direct and indirect impacts from the research of a physics professor. Science and Public Policy, 41(6), 701-719. doi:10.1093/scipol/sct098

[43] Edsand, H.-E., (2017). Identifying barriers to wind energy diffusion in Colombia: A function analysis of the technological innovation system and the wider context. Technology in Society, 49, 1-15.

[44] Dewald, U., Achternbosch, M., (2016). Why more sustainable cements failed so far? Disruptive innovations and their barriers in a basic industry. Environmental Innovation and Societal Transitions, 19, 15-30.

[45] Randelli, F., Rocchi, B., (2017). Analysing the role of consumers within technological innovation systems: The case of alternative food networks. Environmental Innovation and Societal Transitions.

[46] Jansma, S. R., Gosselt, J. F., de Jong, M. D. T., (2017). Technological start-ups in the innovation system: an actor-oriented perspective. Technology Analysis & Strategic Management, 1-13.

[47] Planko, J., Cramer, J., Hekkert, M. P., Chappin, M. M. H., (2017). Combining the technological innovation systems framework with the entrepreneurs' perspective on innovation. Technology Analysis & Strategic Management, 29(6), 614-625.

[48] Sambo, P., Alexander, P., (2018). A scheme of analysis for eVoting as a technological innovation system. Electronic Journal of Information Systems in Developing Countries, 84(2), e12020.

[49] Mäkitie, T., Andersen, A. D., Hanson, J., Normann, H. E., Thune, T. M., (2018). Established sectors expediting clean technology industries? The Norwegian oil and gas sector's influence on offshore wind power. Journal of Cleaner Production, 177, 813-823.

[50] Haley, B., (2017). Designing the public sector to promote sustainability transitions: Institutional principles and a case study of ARPA-E. Environmental Innovation and Societal Transitions, 25, 107-121.

[51] Lukkarinen, J., Berg, A., Salo, M., Tainio, P., Alhola, K., Antikainen, R., (2018). An intermediary approach to technological innovation systems (TIS)—The case of the cleantech sector in Finland. Environmental Innovation and Societal Transitions, 26, 136-146.
۲۰[ Downloaded from jamlu.liau.ac.ir on 2022-06-17 ]